\documentclass[a4paper,fleqn,usenatbib]{mnras}

\usepackage[T1]{fontenc}
\usepackage{ae,aecompl}

\usepackage{graphicx}	
\usepackage{amsmath}	
\usepackage{amssymb}	

\def\msun{{\rm\,M_\odot}}

\def\gtsima{$\; \buildrel > \over \sim \;$}
\def\simgt{\lower.5ex\hbox{\gtsima}}

\def\msun{\hbox{M$_\odot$}}
\def\songmei2022{https://doi.org/10.48550/arXiv.2212.11034}

\title[Multiple populations in star clusters]{A possible trigger of the multiple population phenomenon in star clusters}

\author[Andr\'es E. Piatti]{
Andr\'es E. Piatti$^{1,2}$\thanks{E-mail: andres.piatti@fcen.uncu.edu.ar} \\
$^{1}$Instituto Interdisciplinario de Ciencias B\'asicas (ICB), CONICET-UNCuyo, 
Padre J. Contreras 1300, M5502JMA, Mendoza, Argentina\\
$^{2}$Consejo Nacional de Investigaciones Cient\'{\i}ficas y T\'ecnicas, Godoy Cruz 
2290, C1425FQB,  Buenos Aires, Argentina\\
}

\date{Accepted XXX. Received YYY; in original form ZZZ}

\pubyear{}

\begin{document}
\label{firstpage}
\pagerange{\pageref{firstpage}--\pageref{lastpage}}
\maketitle

\begin{abstract}
Multiple populations (MPs) is a intra-star cluster phenomenon consisting
in star-to-star variation of the abundance of  some light chemical elements.
They have been observed in many star clusters, most of them old globular
clusters, populating the Milky Way and other satellite galaxies. Since the
study of MPs became more systematic, different astrophysical parameters
have been claimed to be the main responsible for its occurrence. However,
at the present time, no attempt would seem to have solved this conundrum.
This work deals with a potential trigger of the MPs phenomenon, based on the
gathered observational evidence of the existence of MPs in some star clusters
and the absence of its in others. We found that star clusters with MPs mostly formed
during time intervals of intense star formation activity in a galaxy, for instance
during the galaxy formation epoch, a close galaxy encounter, etc. 
At those time intervals where relative peaks in the galaxy star formation rate
occur, star clusters with masses above a lower mass limit harbour MPs.
This lower star cluster mass limit would marginally depend on the star cluster age.
\end{abstract}




\section{Introduction}

In a broad context, the multiple populations (MPs) phenomenon refers to the
existence in a star cluster of -at least- two groups of stars which
differentiate one to each other by their distinctive light element abundances
\citep{bl2018}. Since the phenomenon started to be studied systematically, 
there have been different astrophysical features claimed to play an
important role in shaping the MPs.  For instance, \citet{carrettaetal10}
provide one of the earliest systematic and homogeneous analyses, concluding that 
most chemical signatures observed in globular clusters can be attributed primarily 
to a few key parameters, namely: metallicity, mass, and cluster age. \citet{martocchiaetal2018a}
studied nine massive star clusters in the Large Magellanic Cloud (LMC)
from {\it Hubble Space Telescope (HST)} photometry, with ages from $\sim$ 1.5
up to 11.0 Gyr, and found that there is an age dependence among those with 
detected MPs (ages $\ge$ 2.0 Gyr). According to \citet{martocchiaetal2018a} and
reinforced by \citet{martocchiaetal2019}, ancient and young globular clusters 
share a common formation process, which should be reflected in the stellar 
evolutionary models. However, because the existence of old  star clusters with 
single star population has also been confirmed 
\citep{miloneetal2017,miloneetal2018b,lagioiaetal2019,frelijjetal2021,lagioiaetal2025}, the 
suggested dependence on age would 
come out in addition to the dependence on another parameters. On the other hand, 
among Milky Way (MW) open clusters
have not been found trails of MPs \citep[see, e.g.][]{bragagliaetal2024}.

The total star cluster mass has been proposed by \citet{lagioiaetal2019} 
as the main driver of the MPs. They analyzed  $HST$ data of 68 MW and
extragalactic globular clusters to confirm that the width of the
red giant branch in the employed magnitude versus pseudo-colour diagram strongly
correlates with the overall metallicity ([Fe/H] $<$ -0.5 dex) 
\citep[see also][]{miloneetal2017,miloneetal2018b}. When they removed 
the metallicity dependence from the pseudo-colour, they found that it 
shows a strong correlation with the star cluster mass. Hence, they concluded 
that the cluster mass is the main factor affecting the properties of MPs. 
Curiously, the trend of the pseudo-colour with the overall metallicity flattens for 
-1.0 $<$ [Fe/H] $<$ -0.5, and declines its sensibility for more metal-rich 
objects \citep[see, Figures~7 and 8 of][]{lagioiaetal2019}. This trend  may imply
the need to confirm any dependence of the MPs with the cluster mass for star clusters 
more metal-rich than [Fe/H] $\sim$ -0.5 dex.

The study of the formation mechanism of the MPs has led \citet{miloneetal2020} to 
investigate whether the host galaxy has its own role in the light of the
galaxy assembly. Indeed, in the case of the MW, the primordial star population 
or first generation stars have similar chemical composition as that of halo stars with the same 
metallicity, while the secondary star population or second generation stars
show enhancement in the abundance of some chemical elements. They studied
eleven star clusters of both Magellanic Clouds with ages between $\sim$1.5 and
11.0 Gyr from the so-called chromosome map --a two pseudo-colour diagram-- built 
using $HST$ data, and compared them with those of 59 MW globular clusters.
Based on the behaviour of the fraction of first population stars, they found that MPs 
of star clusters do not show any significant different
associated with their respective host galaxies. Later, \citet{vanarajetal2021}
analyzed two additional LMC globular clusters using imaging data from the $HST$
and found that their abundance variations do not distinguish from those of
MW globular clusters. From this outcome they concluded that the galaxy
environment does not play an important role in the formation of MPs 
\citep[see, also][]{leeetal2025}. However, \citet{mm2022} reported that the host 
galaxy may play a role, so that further investigations are needed to assess on
the importance of the effects of the host galaxy on MPs.

Although the star-to-star abundance variation in light elements  --in the
form of N-C and/or Na-O anti-correlation-- has been the flagship evidence of
the existence of MPs \citep{carretta2019,piatti2020a}, variations in some heavy 
elements have also been
found in some globular clusters, and hence used as proxy to assess on the
presence of MPs \citep{carrettaetal2009,pk2018}. However, a general consensus
of heavy element variations among MPs as a key parameter is still needed.
While some MW bulge globular clusters and nuclear star clusters seem to
present abundance variation in heavy elements
\citep{johnsonetal2017,johnsonetal2017b,fernandeztrincadoetal2021,legnardietal2022,schiappaccaseulloaetal2025},
some halo globular clusters and extragalactic star clusters with MPs do not
exhibit any spread \citep{rainetal2019,marinoetal2019,henaoetal2025}. Helium enrichment in 
second generation stars has also been detected \citep{chantereauetal2019,jietal2023,lietal2025},
alongside with variations in the abundance of light elements, so that
helium abundance  would not appear to stand for a major driver of MPs alone either.

Additionally to the different astrophysical properties proposed as a main responsible of
the MPs phenomenon, some studies have appeared recently showing some inconsistencies between 
previous results coming from $HST$ imaging data and high-dispersion spectroscopic studies 
\citep{cb2024,dondoglioetal2025,jangetal2025}. These recent outcomes blur the general knowledge
of the MPs phenomenon, for which it is still pending further investigations to tackle the 
responsible of its occurrence. As far as we are aware, MPs have been seen in star clusters 
with ages
larger than $\sim$ 1.5 Gyr \citep{cadelanoetal2022}, and with masses larger than $\sim$ 
10$^4$ $\msun$
\citep[][]{salgadoetal2022}, and is essentially featured by star-to-star
variations in the abundance of light chemical elements. The MPs phenomenon
would seem to be essentially the same, regardless the galaxy where the
star clusters formed. \citet{lagioiaetal2019} examined more than 40
astrophysical properties of star clusters and found no strong correlation
of the MPs phenomenon with any of them, except the aforementioned link with 
the star cluster mass.

In this work, we introduce for the first time another 
astrophysical quantity that could be considered as a trigger of the MPs
phenomenon. This means that it could mark the environmental conditions
in order to unleash the formation of second generation stars, and hence
the manifestation of the MP phenomenon from observable signatures,
such as light element abundance variation, age, mass, etc.
In what follows, we describe below the justification of a novel approach and 
discuss its scope to explain the appearance of MPs in star clusters, based on the 
available observational evidence.

\section{Analysis and discussion}

The unavoidable questions that arises in the context of the occurrence of the
MPs phenomenon is: what is the cause or starting point that triggers MPs in star 
clusters?  The answer would not seem to be the star-to-star variation of light 
elements, because this feature appears as a consequence of the action of
the cause being searched for. Instead, we are seeking some cause with
the peculiar ability to be correlated with MPs in very different scenarios.
For instance, it is expected that such a cause is correlated with MPs in
ancient star clusters, but with not every old star clusters. In some galaxies, it  
should also point to MPs at moderately old and intermediate ages; the number
of star clusters with MPs would seem to depend on it as well. Although massive 
star clusters have been favourites to find MPs, the expected cause should
also highlight MPs in less massive star clusters at different age regimes.

As far as the dependence on the host galaxy is considered, recent studies
dealt with a variety of star clusters' parameters and showed no dependence
with MPs. However, if the environment (i.e., host galaxy's properties) is meant 
to  be probed as a potential main driver of MPs, the mass of the galaxy, its
star formation rate (SFR) or others galaxy's features would seem to be more
appropriate to infer any correlation between MPs and the environmental
condition in that galaxy. We note that the  cause for MPs should not
be searched for in the internal star clusters' physical conditions, but at a
larger galaxy scale, because MPs are seen in star clusters spatially
distributed and supposedly formed across the whole galaxy extension.
That sought cause should engulf the population of star clusters distributed
throughout the host galaxy to imprint inside them the necessary conditions 
to form MPs.

These ideas are in the line of the interpretation provided by \citet{piatti2020a}
with respect to the Na abundance enhancement in MW globular clusters, which has been the most 
frequent observational evidence to assess the existence of MPs. He suggested that 
the difference in the Na abundance content between the second and first generation 
stars is related to a characteristic called 'cosmological vitality', which
refers to the powerful strength deployed at the epoch of the Universe formation, that 
soon after became a more quiescent nucleosynthesis activity. Although MPs are
thought to be formed within the star clusters from intra-cluster processes, 
such as interactions between stars, they are triggered by that cosmological vitality.

The approach chosen in this work in order to uncover some overall cause 
correlated with MPs is necessarily limited to the available observational information 
on the existence of MPs in star clusters. Therefore, further investigations of
this phenomenon from a larger sample of star clusters will provide more
definitive trends to this respect. Table~\ref{tab1} (see Appendix) lists the star clusters' 
properties gathered from the literature employed in this study. We limited the analysis to
galaxies where both kind of star clusters formed, i.e., with and without MPs. The list of 
star clusters associated to Sagittarius and Helmi streams progenitors, i.e., star
clusters formed in dwarf galaxies that later merged into the MW, as well as those
formed in the MW were taken from \citet{callinghametal2022}. 

The ages and metallicities of globular clusters associated with Sagittarius, the Helmi 
streams and those formed in the MW were taken from \citet{kruijssenetal2019}, except 
for NGC~6229 which is not 
included in their compilation \citep[see, instead,][]{borissovaetal1999}. The masses of these 
globular clusters were taken from the catalog of 
\citet{baumgardtetal2023}\footnote{https://people.smp.uq.edu.au/HolgerBaumgardt/globular/}. 
Open clusters' ages and metallicites
were taken from the works about the detection of MPs (see last column in Table~\ref{tab1}), 
and their masses from \citet{hr2024}. For LMC globular clusters, we
used the ages, metallicities and masses compiled in \citet{pm2018} and \citet{piattietal2019},
respectively, and for younger LMC star clusters we adopted the values given in the studies 
of their MPs (see references in Table~\ref{tab1}). In the case of the
Small Magellanic Cloud (SMC) star clusters, their ages and metallicities come from
\citet{piatti2023c}, and their
masses come from the works of \citet{mvdm05} and \citet{getal11}, or from the references
of their MPs studies (Lindsay~8 and 113). The star clusters are ordered according to decreasing 
ages. The last columns of Table~\ref{tab1} point out whether MPs have been detected, alongside the 
corresponding references.

We first examined Table~\ref{tab1} and glimpsed that the oldest star clusters usually
exhibit MPs, coinciding with one of the time intervals where the galaxies
deployed an intense star formation activity. In the Magellanic Clouds younger
star clusters also harbour MPs, and curiously, their ages seem to match the
known periods of mutual interaction or interaction with the MW, so that
relative important enhanced star formation events could take place. Indeed,
around 2-3 Gyr both Magellanic Clouds experienced bursting star cluster
formation episodes \citep{p11a,p11b,pg13}, while the SMC would seem to have 
witnessed another one at $\sim$ 6-8 Gyr \citep{richetal2000,tb2009}.
From this observational evidence, we searched for the star formation rates
(SFRs) of each studied galaxy, with the aim of probing any synchronicity
between the periods of more intense star formation and the ages of the star
clusters with MPs, and role reversal.
We assumed that the formation of field stars and star clusters have occurred
concurrently, so that the star cluster frequency - the number of clusters per 
time unit as a function of age - is proportional to the respective SFR
\citep{p14b,piatti2021e}. The adopted SFRs are the following: MW 
\citep{ruizlaraetal2020,spitonietal2024}; Sagittarius \citep{deboeretal2015}; 
Helmi streams progenitor \citep{ruizlaraetal2022}, LMC and SMC \citep{massanaetal2022},
respectively.

\begin{figure}
\includegraphics[width=\columnwidth]{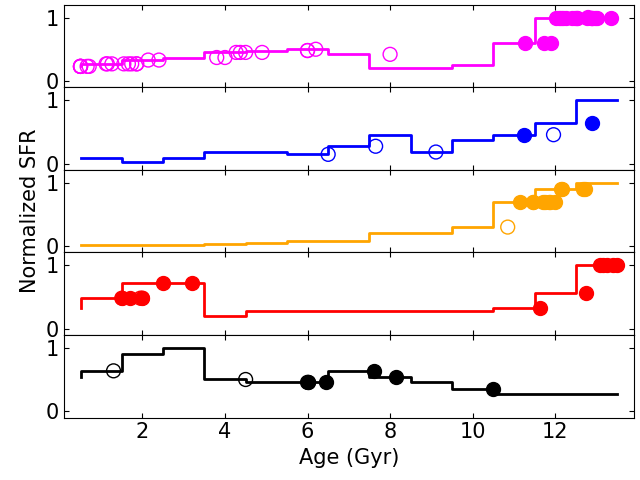}
\caption{SFRs of the MW \citep[magenta,][]{ruizlaraetal2020,spitonietal2024}; Sagittarius
\citep[blue,][]{deboeretal2015}; Helmi streams \citep[orange,][]{ruizlaraetal2022}; LMC
 and SMC \citep[red and black,][]{massanaetal2022} are shown, respectively. Open and
filled circles are star clusters without and with MPs, respectively, drawn on the
SFRs at their corresponding ages.}
\label{fig1}
\end{figure}

Figure~\ref{fig1} depicts the normalized SFRs for the studied galaxies.
All the galaxies, except the SMC, shows an intense star formation activity
at ages $>$ 12 Gyr, a period of time that coincides with the ages of a vast
majority of globular clusters with MPs. The bursting star formation period 
in the LMC is also populated by star clusters with MPs, as well as the early
enhanced star formation epoch in the SMC. Conversely, outside these periods
of time characterized by an intense star formation activity, star clusters
would not seem to show MPs. This simple diagnostic picture suggests that the
powerful strength deployed during periods of relatively high SFRs
can form star clusters and trigger inside them processes 
(e.g., stellar interaction, stellar evolution, nucleosynthesis, etc) that
lead to the formation of second generation stars.  Note that this scenario
of intense star formation activity is compatible  with the formation also of  star 
clusters without MPs (e.g., Sagittarius, Helmi streams),
Nevertheless, intense SFR events would seem
to be privileged scenarios for triggering  MP in star clusters.

An examination of Table~\ref{tab1} reveals that star clusters with MPs
have notably larger masses than star clusters without MPs, which rises
the possibility that there could be a connection between enhanced SFR periods and
the formation of more massive star clusters. According to the integrated 
cloud-wide initial mass  function theory, the mass of the most massive star cluster 
is positively proportional to the SFR, the SFR surface density of the cloud where it formed, 
as well as, to the mass of the cloud and  the column density \citep{zhouetal2025}.
Likewise, \citet{lietal2018} carried out simulations  and found that the fraction of 
clustered star formation and maximum cluster mass increase with the SFR surface 
density. Interestingly, \citet{renzinietal2022} analysed the formation of
globular clusters in an overcooling scenario and noted that a lower mass limit is
needed in order to form globular clusters with MP. \citet{bereketal2023} performed 
logistic regressions using the SFR and 
the total stellar mass in the galaxy as predictors, and found that the SFR is the better 
predictor for the probability of hosting clusters and the total mass in the cluster system. 
When they compared their results to similar models for old globular clusters, they
concluded that the star cluster formation was more abundant and more efficient at 
higher redshifts, likely because of the high gas content of galaxies at that time.

\begin{figure}
\includegraphics[width=\columnwidth]{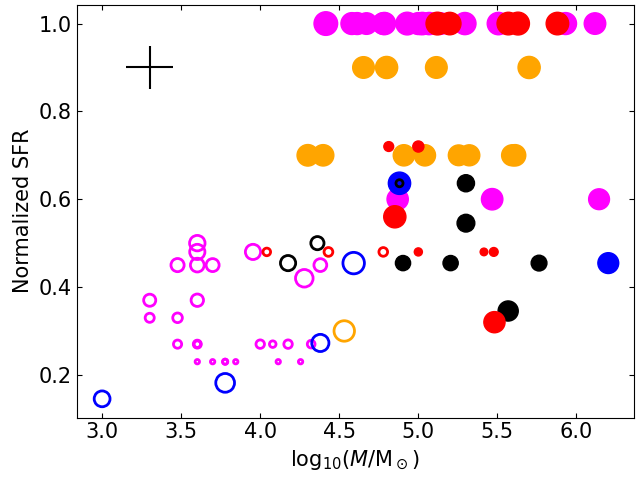}
\caption{Normalized SFR versus star cluster mass. The symbols are
as in Figure~\ref{fig1} with sizes proportional to the star cluster ages. 
Typical error bars are also shown.}
\label{fig2}
\end{figure}

We explored the star cluster mass versus the normalized SFR (see 
Figure~\ref{fig2}) and found a broad correspondence, in the sense that
the larger the mass, the larger the normalized SFR. Since intense
SFR events are privileged scenarios for the formation of massive star
clusters, which in turn are found more frequently harbouring MPs, Figure~\ref{fig2}
reveals for the first time that enhanced SFR periods can more frequently 
provide the necessary
conditions for massive star cluster formation, which in turn initiate MPs.  
Therefore, intense SFR episodes would play as possible triggers of the MP
phenomenon, which is manifested (observed) in more massive star clusters.
Note that we considered present-day star cluster masses.
Star clusters  older than $\sim$ 4 Gyr can lose nearly 50$\%$ of their
initial masses due to stellar evolution \citep{lamersetal2005a}, and $\sim$ 5-45$\%$
due to relaxation and tidal effects \citep{piattietal2019}. However, star
clusters disrupted would have had on average initial masses smaller than those for 
harbouring MPs, since there is a narrow lower mass limit for having MPs
that marginally depends on age, as follows:

\begin{equation}
{\rm log}_{\rm 10}(M/M_{\odot}) = (-0.036\pm0.005)\times (t/{\rm Gyr}) 
+ 4.916\pm0.100
\end{equation}

\noindent where $M$ and $t$ are the cluster mass and age, respectively.

Figures~\ref{fig1} and \ref{fig2} lead to conclude that MPs are seen in more
massive star clusters, which preferentially formed during enhanced SFR periods.
The lower mass limit (Eq.~(1)) would also seem to explain 
the existence of star clusters without MPs formed during periods of intense star 
formation. In the context of Figures~\ref{fig1} and \ref{fig2}, the MW globular clusters
formed in-situ (magenta coloured symbols) harbour MPs, with the exception of
E~3. NGC~6717 is of a similar age and mass to E~3, and shows MPs. We
recall that the SFR can vary with the position in the galaxy \citep{p14b,deboeretal2015}, 
so that the massive star clusters formed too.
MW open clusters do not show MPs because of their masses, regardless they
formed in periods of moderately high SFRs. In Sagittarius, Arp~2 is at the
cluster mass limit to have MPs, while the other younger  globular clusters
are indeed still less massive. None of them have MPs. Ruprecht~106 in the Helmi
streams does not have MPs either because of its mass or because its formed
at the end of an intense star formation epoch. SMC star clusters with MPs
have the enough mass and formed during a bursting star formation event
\citep{tb2009,p12a}, except NGC~121, whose relatively more metal-rich
chemical composition at it age (10.5 Gyr) could imply an SMC ex-situ
origin.

In order to quantify the role played by intense SFR events and massive star clusters
in the existence of MPs, we performed a multivariate logistic regression from
Figure~\ref{fig2}. We modeled binary outcomes (MPs or no MPs) by using
\texttt{scikit-learn}\footnote{https://scikit-learn.org/stable/}. We
split the data into training and testing sets, trained the logistic
regression model, made the predictions, and uncovered the impact on
the MP phenomenon of the normalized SFR and the star cluster mass. 
When using simultaneously the normalized SFR and the star cluster mass as independent 
predictors, we obtained coefficients of 6.60$\pm$3.10 and 5.93$\pm$2.14, and p-values
of 0.03 and 0.00, respectively, with a goodness of the model fit of $R^2$ = 0.83, and 
an accuracy of the logistic regression of 94$\%$.  This means that both parameters
have a similar importance in shaping MPs.  If we used the normalized SFR or the star
cluster mass alone as a predictor, then the respective coefficients would turn to be
13.71$\pm$3.52 (fit accuracy = 94$\%$), and 7.55$\pm$2.40 (fit accuracy = 88$\%$),
which implies that the normalized SFR is a more important predictor than the
star cluster mass when playing as the only driver of MPs. Note, however, that
an intense SFR is a prerequisite to form massive star clusters.

\section{Conclusions}

The MPs phenomenon has been observed in star clusters spanning a wide
range of age and mass, among other astrophysical parameters. The
puzzle of correlating these various star cluster properties has led
to seek for different main drivers of such a phenomenon with relative
success. Here we introduce a potential MPs trigger based on the
gathered observational evidence of its existence in some star clusters
and its absence in others. We propose that star clusters are prone to
harbour MPs if they formed during time intervals of the galaxy lifetime
during which a relative peak of its SFR occurs. The time intervals with relatively 
high SFRs can be those at the early epoch of the formation of the Universe 
and those at the close interaction of galaxies and galaxy mergers. 
Star clusters with ages similar to the age of these intense formation
events can form second generation stars. Nevertheless, the lookback time
of the star cluster formation alone does not fully explain the MPs phenomenon.
Seemingly, there is a lower star cluster mass limit, below which the
intra-cluster conditions would not favour the formation of second generation
stars. We found that such a lower star cluster mass limit does marginally depend
on the star cluster age. 

We note that an intense SFR is not a driver of MPs, as it is the case of the
cluster mass, the variation in light element abundances, etc, which manifest the
existence of MPs. Instead, an intense SFR would seem to be the cause, the 
starting point, from which certain astrophysical properties witness the existence of
MPs. As far as we are aware, there is no results available in the literature
related to SFRs suggesting any connection with MPs, nor any studies of MPs in massive
star clusters linking the MPs to intense SFRs either. Although the relation between
intense SFRs and MPs is not unexpected, this is the first time its relation
 is introduced (Figure~\ref{fig2}) and quantitatively assessed. The present outcomes
open a variety of new possibilities in the study of MPs. For instance, it is more
straightforward to know the periods of intense SFR in a galaxy where to seek for star
clusters with MPs than to hit the target from a blind search of them. This implies 
that the completeness of the population of star clusters with MPs in a galaxy 
could be statistically achieved more robustly.

\section*{Acknowledgements}
We thank the referee for the thorough reading of the manuscript and
timely suggestions to improve it. 

\section{Data availability}

Data used in this work are available upon request to the author.




\begin{thebibliography}{}
\makeatletter
\relax
\def\mn@urlcharsother{\let\do\@makeother \do\$\do\&\do\#\do\^\do\_\do\%\do\~}
\def\mn@doi{\begingroup\mn@urlcharsother \@ifnextchar [ {\mn@doi@}
  {\mn@doi@[]}}
\def\mn@doi@[#1]#2{\def\@tempa{#1}\ifx\@tempa\@empty \href
  {http://dx.doi.org/#2} {doi:#2}\else \href {http://dx.doi.org/#2} {#1}\fi
  \endgroup}
\def\mn@eprint#1#2{\mn@eprint@#1:#2::\@nil}
\def\mn@eprint@arXiv#1{\href {http://arxiv.org/abs/#1} {{\tt arXiv:#1}}}
\def\mn@eprint@dblp#1{\href {http://dblp.uni-trier.de/rec/bibtex/#1.xml}
  {dblp:#1}}
\def\mn@eprint@#1:#2:#3:#4\@nil{\def\@tempa {#1}\def\@tempb {#2}\def\@tempc
  {#3}\ifx \@tempc \@empty \let \@tempc \@tempb \let \@tempb \@tempa \fi \ifx
  \@tempb \@empty \def\@tempb {arXiv}\fi \@ifundefined
  {mn@eprint@\@tempb}{\@tempb:\@tempc}{\expandafter \expandafter \csname
  mn@eprint@\@tempb\endcsname \expandafter{\@tempc}}}

\bibitem[\protect\citeauthoryear{{Bastian} \& {Lardo}}{{Bastian} \&
  {Lardo}}{2018}]{bl2018}
{Bastian} N.,  {Lardo} C.,  2018, \mn@doi [\araa]
  {10.1146/annurev-astro-081817-051839}, \href
  {https://ui.adsabs.harvard.edu/abs/2018ARA&A..56...83B} {56, 83}

\bibitem[\protect\citeauthoryear{{Baumgardt}, {H{\'e}nault-Brunet}, {Dickson}
  \& {Sollima}}{{Baumgardt} et~al.}{2023}]{baumgardtetal2023}
{Baumgardt} H.,  {H{\'e}nault-Brunet} V.,  {Dickson} N.,   {Sollima} A.,  2023,
  \mn@doi [\mnras] {10.1093/mnras/stad631}, \href
  {https://ui.adsabs.harvard.edu/abs/2023MNRAS.521.3991B} {521, 3991}

\bibitem[\protect\citeauthoryear{{Berek}, {Reina-Campos}, {Eadie}  \&
  {Sills}}{{Berek} et~al.}{2023}]{bereketal2023}
{Berek} S.~C.,  {Reina-Campos} M.,  {Eadie} G.,   {Sills} A.,  2023, \mn@doi
  [\mnras] {10.1093/mnras/stad2302}, \href
  {https://ui.adsabs.harvard.edu/abs/2023MNRAS.525.1902B} {525, 1902}

\bibitem[\protect\citeauthoryear{{Borissova}, {Catelan}, {Ferraro}, {Spassova},
  {Buonanno}, {Iannicola}, {Richtler}  \& {Sweigart}}{{Borissova}
  et~al.}{1999}]{borissovaetal1999}
{Borissova} J.,  {Catelan} M.,  {Ferraro} F.~R.,  {Spassova} N.,  {Buonanno}
  R.,  {Iannicola} G.,  {Richtler} T.,   {Sweigart} A.~V.,  1999, \mn@doi
  [\aap] {10.48550/arXiv.astro-ph/9810355}, \href
  {https://ui.adsabs.harvard.edu/abs/1999A&A...343..813B} {343, 813}

\bibitem[\protect\citeauthoryear{{Bragaglia} et~al.,}{{Bragaglia}
  et~al.}{2024}]{bragagliaetal2024}
{Bragaglia} A.,  et~al., 2024, \mn@doi [\aap] {10.1051/0004-6361/202245144},
  \href {https://ui.adsabs.harvard.edu/abs/2024A&A...687A.124B} {687, A124}

\bibitem[\protect\citeauthoryear{{Cadelano}, {Dalessandro}, {Salaris},
  {Bastian}, {Mucciarelli}, {Saracino}, {Martocchia}  \&
  {Cabrera-Ziri}}{{Cadelano} et~al.}{2022}]{cadelanoetal2022}
{Cadelano} M.,  {Dalessandro} E.,  {Salaris} M.,  {Bastian} N.,  {Mucciarelli}
  A.,  {Saracino} S.,  {Martocchia} S.,   {Cabrera-Ziri} I.,  2022, \mn@doi
  [\apjl] {10.3847/2041-8213/ac424a}, \href
  {https://ui.adsabs.harvard.edu/abs/2022ApJ...924L...2C} {924, L2}

\bibitem[\protect\citeauthoryear{{Callingham}, {Cautun}, {Deason}, {Frenk},
  {Grand}  \& {Marinacci}}{{Callingham} et~al.}{2022}]{callinghametal2022}
{Callingham} T.~M.,  {Cautun} M.,  {Deason} A.~J.,  {Frenk} C.~S.,  {Grand} R.
  J.~J.,   {Marinacci} F.,  2022, \mn@doi [\mnras] {10.1093/mnras/stac1145},
  \href {https://ui.adsabs.harvard.edu/abs/2022MNRAS.513.4107C} {513, 4107}

\bibitem[\protect\citeauthoryear{{Carretta}}{{Carretta}}{2019}]{carretta2019}
{Carretta} E.,  2019, \mn@doi [\aap] {10.1051/0004-6361/201935110}, \href
  {https://ui.adsabs.harvard.edu/abs/2019A&A...624A..24C} {624, A24}

\bibitem[\protect\citeauthoryear{{Carretta} \& {Bragaglia}}{{Carretta} \&
  {Bragaglia}}{2021}]{cb2021}
{Carretta} E.,  {Bragaglia} A.,  2021, \mn@doi [\aap]
  {10.1051/0004-6361/202039392}, \href
  {https://ui.adsabs.harvard.edu/abs/2021A&A...646A...9C} {646, A9}

\bibitem[\protect\citeauthoryear{{Carretta} \& {Bragaglia}}{{Carretta} \&
  {Bragaglia}}{2025}]{cb2024}
{Carretta} E.,  {Bragaglia} A.,  2025, \mn@doi [\aap]
  {10.1051/0004-6361/202553755}, \href
  {https://ui.adsabs.harvard.edu/abs/2025A&A...696A.120C} {696, A120}

\bibitem[\protect\citeauthoryear{{Carretta}, {Bragaglia}, {Gratton}, {D'Orazi}
  \& {Lucatello}}{{Carretta} et~al.}{2009}]{carrettaetal2009}
{Carretta} E.,  {Bragaglia} A.,  {Gratton} R.,  {D'Orazi} V.,   {Lucatello} S.,
   2009, \mn@doi [\aap] {10.1051/0004-6361/200913003}, \href
  {http://adsabs.harvard.edu/abs/2009A%26A...508..695C} {508, 695}

\bibitem[\protect\citeauthoryear{{Carretta}, {Bragaglia}, {Gratton},
  {Recio-Blanco}, {Lucatello}, {D'Orazi}  \& {Cassisi}}{{Carretta}
  et~al.}{2010}]{carrettaetal10}
{Carretta} E.,  {Bragaglia} A.,  {Gratton} R.~G.,  {Recio-Blanco} A.,
  {Lucatello} S.,  {D'Orazi} V.,   {Cassisi} S.,  2010, \mn@doi [\aap]
  {10.1051/0004-6361/200913451}, 516, A55

\bibitem[\protect\citeauthoryear{{Chantereau}, {Salaris}, {Bastian}  \&
  {Martocchia}}{{Chantereau} et~al.}{2019}]{chantereauetal2019}
{Chantereau} W.,  {Salaris} M.,  {Bastian} N.,   {Martocchia} S.,  2019,
  \mn@doi [\mnras] {10.1093/mnras/stz378}, \href
  {https://ui.adsabs.harvard.edu/abs/2019MNRAS.484.5236C} {484, 5236}

\bibitem[\protect\citeauthoryear{{Dondoglio} et~al.,}{{Dondoglio}
  et~al.}{2025}]{dondoglioetal2025}
{Dondoglio} E.,  et~al., 2025, \mn@doi [\aap] {10.1051/0004-6361/202453024},
  \href {https://ui.adsabs.harvard.edu/abs/2025A&A...697A.135D} {697, A135}

\bibitem[\protect\citeauthoryear{{Fern{\'a}ndez-Trincado}
  et~al.,}{{Fern{\'a}ndez-Trincado} et~al.}{2021}]{fernandeztrincadoetal2021}
{Fern{\'a}ndez-Trincado} J.~G.,  et~al., 2021, \mn@doi [\apjl]
  {10.3847/2041-8213/ac1c7e}, \href
  {https://ui.adsabs.harvard.edu/abs/2021ApJ...918L...9F} {918, L9}

\bibitem[\protect\citeauthoryear{{Frelijj}, {Villanova}, {Mu{\~n}oz}  \&
  {Fern{\'a}ndez-Trincado}}{{Frelijj} et~al.}{2021}]{frelijjetal2021}
{Frelijj} H.,  {Villanova} S.,  {Mu{\~n}oz} C.,   {Fern{\'a}ndez-Trincado}
  J.~G.,  2021, \mn@doi [\mnras] {10.1093/mnras/stab443}, \href
  {https://ui.adsabs.harvard.edu/abs/2021MNRAS.503..867F} {503, 867}

\bibitem[\protect\citeauthoryear{{Gilligan} et~al.,}{{Gilligan}
  et~al.}{2020}]{gilliganetal2020}
{Gilligan} C.~K.,  et~al., 2020, \mn@doi [\mnras] {10.1093/mnras/staa822},
  \href {https://ui.adsabs.harvard.edu/abs/2020MNRAS.494.1946G} {494, 1946}

\bibitem[\protect\citeauthoryear{{Glatt} et~al.,}{{Glatt}
  et~al.}{2011}]{getal11}
{Glatt} K.,  et~al., 2011, \mn@doi [\aj] {10.1088/0004-6256/142/2/36}, 142, 36

\bibitem[\protect\citeauthoryear{{Henao}, {Villanova}, {Geisler}  \&
  {Fern{\'a}ndez-Trincado}}{{Henao} et~al.}{2025}]{henaoetal2025}
{Henao} L.,  {Villanova} S.,  {Geisler} D.,   {Fern{\'a}ndez-Trincado} J.~G.,
  2025, \mn@doi [\aap] {10.1051/0004-6361/202451793}, \href
  {https://ui.adsabs.harvard.edu/abs/2025A&A...696A.154H} {696, A154}

\bibitem[\protect\citeauthoryear{{Huang} et~al.,}{{Huang}
  et~al.}{2024}]{huangetal2024}
{Huang} R.,  et~al., 2024, \mn@doi [Science China Physics, Mechanics, and
  Astronomy] {10.1007/s11433-023-2332-5}, \href
  {https://ui.adsabs.harvard.edu/abs/2024SCPMA..6759513H} {67, 259513}

\bibitem[\protect\citeauthoryear{{Hunt} \& {Reffert}}{{Hunt} \&
  {Reffert}}{2024}]{hr2024}
{Hunt} E.~L.,  {Reffert} S.,  2024, \mn@doi [\aap]
  {10.1051/0004-6361/202348662}, \href
  {https://ui.adsabs.harvard.edu/abs/2024A&A...686A..42H} {686, A42}

\bibitem[\protect\citeauthoryear{{Jang} et~al.,}{{Jang}
  et~al.}{2025}]{jangetal2025}
{Jang} S.,  et~al., 2025, \mn@doi [\apj] {10.3847/1538-4357/adafa1}, \href
  {https://ui.adsabs.harvard.edu/abs/2025ApJ...981...57J} {981, 57}

\bibitem[\protect\citeauthoryear{{Ji}, {Deng}, {Chen}, {Li}  \& {Liu}}{{Ji}
  et~al.}{2023}]{jietal2023}
{Ji} X.,  {Deng} L.,  {Chen} Y.,  {Li} C.,   {Liu} C.,  2023, \mn@doi [Research
  in Astronomy and Astrophysics] {10.1088/1674-4527/accdc4}, \href
  {https://ui.adsabs.harvard.edu/abs/2023RAA....23g5009J} {23, 075009}

\bibitem[\protect\citeauthoryear{{Johnson}, {Caldwell}, {Rich}  \&
  {Walker}}{{Johnson} et~al.}{2017a}]{johnsonetal2017b}
{Johnson} C.~I.,  {Caldwell} N.,  {Rich} R.~M.,   {Walker} M.~G.,  2017a,
  \mn@doi [\aj] {10.3847/1538-3881/aa86ac}, \href
  {https://ui.adsabs.harvard.edu/abs/2017AJ....154..155J} {154, 155}

\bibitem[\protect\citeauthoryear{{Johnson}, {Caldwell}, {Rich}, {Mateo},
  {Bailey}, {Clarkson}, {Olszewski}  \& {Walker}}{{Johnson}
  et~al.}{2017b}]{johnsonetal2017}
{Johnson} C.~I.,  {Caldwell} N.,  {Rich} R.~M.,  {Mateo} M.,  {Bailey} III
  J.~I.,  {Clarkson} W.~I.,  {Olszewski} E.~W.,   {Walker} M.~G.,  2017b,
  \mn@doi [\apj] {10.3847/1538-4357/836/2/168}, \href
  {https://ui.adsabs.harvard.edu/abs/2017ApJ...836..168J} {836, 168}

\bibitem[\protect\citeauthoryear{{Kapse}, {de Grijs}, {Kamath}  \&
  {Zucker}}{{Kapse} et~al.}{2022}]{kapseetal2022}
{Kapse} S.,  {de Grijs} R.,  {Kamath} D.,   {Zucker} D.~B.,  2022, \mn@doi
  [\apjl] {10.3847/2041-8213/ac551a}, \href
  {https://ui.adsabs.harvard.edu/abs/2022ApJ...927L..10K} {927, L10}

\bibitem[\protect\citeauthoryear{{Kruijssen}, {Pfeffer}, {Reina-Campos},
  {Crain}  \& {Bastian}}{{Kruijssen} et~al.}{2019}]{kruijssenetal2019}
{Kruijssen} J.~M.~D.,  {Pfeffer} J.~L.,  {Reina-Campos} M.,  {Crain} R.~A.,
  {Bastian} N.,  2019, \mn@doi [\mnras] {10.1093/mnras/sty1609}, \href
  {https://ui.adsabs.harvard.edu/abs/2019MNRAS.486.3180K} {486, 3180}

\bibitem[\protect\citeauthoryear{{Lagioia}, {Milone}, {Marino}, {Cordoni}  \&
  {Tailo}}{{Lagioia} et~al.}{2019}]{lagioiaetal2019}
{Lagioia} E.~P.,  {Milone} A.~P.,  {Marino} A.~F.,  {Cordoni} G.,   {Tailo} M.,
   2019, \mn@doi [\aj] {10.3847/1538-3881/ab45f2}, \href
  {https://ui.adsabs.harvard.edu/abs/2019AJ....158..202L} {158, 202}

\bibitem[\protect\citeauthoryear{{Lagioia} et~al.,}{{Lagioia}
  et~al.}{2025}]{lagioiaetal2025}
{Lagioia} E.~P.,  et~al., 2025, \mn@doi [\apj] {10.3847/1538-4357/ad98ee},
  \href {https://ui.adsabs.harvard.edu/abs/2025ApJ...979...30L} {979, 30}

\bibitem[\protect\citeauthoryear{{Lamers}, {Gieles}, {Bastian}, {Baumgardt},
  {Kharchenko}  \& {Portegies Zwart}}{{Lamers} et~al.}{2005}]{lamersetal2005a}
{Lamers} H.~J.~G.~L.~M.,  {Gieles} M.,  {Bastian} N.,  {Baumgardt} H.,
  {Kharchenko} N.~V.,   {Portegies Zwart} S.,  2005, \mn@doi [\aap]
  {10.1051/0004-6361:20042241}, \href
  {http://adsabs.harvard.edu/abs/2005A%26A...441..117L} {441, 117}

\bibitem[\protect\citeauthoryear{{Lee}, {Kim}, {Kim}, {Sung}, {Kim}  \& {Di
  Mille}}{{Lee} et~al.}{2025}]{leeetal2025}
{Lee} J.-W.,  {Kim} T.-H.,  {Kim} H.-S.,  {Sung} H.-I.,  {Kim} H.,   {Di Mille}
  F.,  2025, \mn@doi [\aj] {10.3847/1538-3881/ada94e}, \href
  {https://ui.adsabs.harvard.edu/abs/2025AJ....169..143L} {169, 143}

\bibitem[\protect\citeauthoryear{{Legnardi} et~al.,}{{Legnardi}
  et~al.}{2022}]{legnardietal2022}
{Legnardi} M.~V.,  et~al., 2022, \mn@doi [\mnras] {10.1093/mnras/stac734},
  \href {https://ui.adsabs.harvard.edu/abs/2022MNRAS.513..735L} {513, 735}

\bibitem[\protect\citeauthoryear{{Li}, {Gnedin}  \& {Gnedin}}{{Li}
  et~al.}{2018}]{lietal2018}
{Li} H.,  {Gnedin} O.~Y.,   {Gnedin} N.~Y.,  2018, \mn@doi [\apj]
  {10.3847/1538-4357/aac9b8}, \href
  {https://ui.adsabs.harvard.edu/abs/2018ApJ...861..107L} {861, 107}

\bibitem[\protect\citeauthoryear{{Li}, {Wang}  \& {Milone}}{{Li}
  et~al.}{2019}]{lietal2019b}
{Li} C.,  {Wang} Y.,   {Milone} A.~P.,  2019, \mn@doi [\apj]
  {10.3847/1538-4357/ab3c54}, \href
  {https://ui.adsabs.harvard.edu/abs/2019ApJ...884...17L} {884, 17}

\bibitem[\protect\citeauthoryear{{Li}, {Wang}, {Ji}  \& {Baumgardt}}{{Li}
  et~al.}{2025}]{lietal2025}
{Li} C.,  {Wang} L.,  {Ji} X.,   {Baumgardt} H.,  2025, \mn@doi [\aap]
  {10.1051/0004-6361/202452952}, \href
  {https://ui.adsabs.harvard.edu/abs/2025A&A...695A..63L} {695, A63}

\bibitem[\protect\citeauthoryear{{Marino} et~al.,}{{Marino}
  et~al.}{2019}]{marinoetal2019}
{Marino} A.~F.,  et~al., 2019, \mn@doi [\mnras] {10.1093/mnras/stz1415}, \href
  {https://ui.adsabs.harvard.edu/abs/2019MNRAS.487.3815M} {487, 3815}

\bibitem[\protect\citeauthoryear{{Martocchia} et~al.,}{{Martocchia}
  et~al.}{2018}]{martocchiaetal2018a}
{Martocchia} S.,  et~al., 2018, \mn@doi [\mnras] {10.1093/mnras/stx2556}, \href
  {http://adsabs.harvard.edu/abs/2018MNRAS.473.2688M} {473, 2688}

\bibitem[\protect\citeauthoryear{{Martocchia} et~al.,}{{Martocchia}
  et~al.}{2019}]{martocchiaetal2019}
{Martocchia} S.,  et~al., 2019, \mn@doi [\mnras] {10.1093/mnras/stz1596}, \href
  {https://ui.adsabs.harvard.edu/abs/2019MNRAS.487.5324M} {487, 5324}

\bibitem[\protect\citeauthoryear{{Massana} et~al.,}{{Massana}
  et~al.}{2022}]{massanaetal2022}
{Massana} P.,  et~al., 2022, \mn@doi [\mnras] {10.1093/mnrasl/slac030}, \href
  {https://ui.adsabs.harvard.edu/abs/2022MNRAS.513L..40M} {513, L40}

\bibitem[\protect\citeauthoryear{{McLaughlin} \& {van der Marel}}{{McLaughlin}
  \& {van der Marel}}{2005}]{mvdm05}
{McLaughlin} D.~E.,  {van der Marel} R.~P.,  2005, \mn@doi [\apjs]
  {10.1086/497429}, \href {http://adsabs.harvard.edu/abs/2005ApJS..161..304M}
  {161, 304}

\bibitem[\protect\citeauthoryear{{Milone} \& {Marino}}{{Milone} \&
  {Marino}}{2022}]{mm2022}
{Milone} A.~P.,  {Marino} A.~F.,  2022, \mn@doi [Universe]
  {10.3390/universe8070359}, \href
  {https://ui.adsabs.harvard.edu/abs/2022Univ....8..359M} {8, 359}

\bibitem[\protect\citeauthoryear{{Milone} et~al.,}{{Milone}
  et~al.}{2017}]{miloneetal2017}
{Milone} A.~P.,  et~al., 2017, \mn@doi [\mnras] {10.1093/mnras/stw2531}, \href
  {https://ui.adsabs.harvard.edu/abs/2017MNRAS.464.3636M} {464, 3636}

\bibitem[\protect\citeauthoryear{{Milone} et~al.,}{{Milone}
  et~al.}{2018}]{miloneetal2018b}
{Milone} A.~P.,  et~al., 2018, \mn@doi [\mnras] {10.1093/mnras/sty2573}, \href
  {https://ui.adsabs.harvard.edu/abs/2018MNRAS.481.5098M} {481, 5098}

\bibitem[\protect\citeauthoryear{{Milone} et~al.,}{{Milone}
  et~al.}{2020}]{miloneetal2020}
{Milone} A.~P.,  et~al., 2020, \mn@doi [\mnras] {10.1093/mnras/stz2999}, \href
  {https://ui.adsabs.harvard.edu/abs/2020MNRAS.491..515M} {491, 515}

\bibitem[\protect\citeauthoryear{{Niederhofer} et~al.,}{{Niederhofer}
  et~al.}{2017a}]{niederhoferetal2017a}
{Niederhofer} F.,  et~al., 2017a, \mn@doi [\mnras] {10.1093/mnras/stw2269},
  \href {https://ui.adsabs.harvard.edu/abs/2017MNRAS.464...94N} {464, 94}

\bibitem[\protect\citeauthoryear{{Niederhofer} et~al.,}{{Niederhofer}
  et~al.}{2017b}]{niederhoferetal2017b}
{Niederhofer} F.,  et~al., 2017b, \mn@doi [\mnras] {10.1093/mnras/stw3084},
  \href {http://adsabs.harvard.edu/abs/2017MNRAS.465.4159N} {465, 4159}

\bibitem[\protect\citeauthoryear{{Oh}, {Nordlander}, {Da Costa}  \&
  {Mackey}}{{Oh} et~al.}{2023}]{ohetal2023}
{Oh} W.~S.,  {Nordlander} T.,  {Da Costa} G.~S.,   {Mackey} A.~D.,  2023,
  \mn@doi [\mnras] {10.1093/mnras/stac3552}, \href
  {https://ui.adsabs.harvard.edu/abs/2023MNRAS.519..831O} {519, 831}

\bibitem[\protect\citeauthoryear{{Pessev}, {Goudfrooij}, {Puzia}  \&
  {Chandar}}{{Pessev} et~al.}{2008}]{pessevetal2008}
{Pessev} P.~M.,  {Goudfrooij} P.,  {Puzia} T.~H.,   {Chandar} R.,  2008,
  \mn@doi [\mnras] {10.1111/j.1365-2966.2008.12935.x}, \href
  {https://ui.adsabs.harvard.edu/abs/2008MNRAS.385.1535P} {385, 1535}

\bibitem[\protect\citeauthoryear{{Piatti}}{{Piatti}}{2011a}]{p11a}
{Piatti} A.~E.,  2011a, \mn@doi [\mnras] {10.1111/j.1745-3933.2011.01139.x},
  418, L40

\bibitem[\protect\citeauthoryear{{Piatti}}{{Piatti}}{2011b}]{p11b}
{Piatti} A.~E.,  2011b, \mn@doi [\mnras] {10.1111/j.1745-3933.2011.01145.x},
  418, L69

\bibitem[\protect\citeauthoryear{{Piatti}}{{Piatti}}{2012}]{p12a}
{Piatti} A.~E.,  2012, \mn@doi [\mnras] {10.1111/j.1365-2966.2012.20684.x},
  422, 1109

\bibitem[\protect\citeauthoryear{{Piatti}}{{Piatti}}{2014}]{p14b}
{Piatti} A.~E.,  2014, \mn@doi [\mnras] {10.1093/mnras/stt1998}, 437, 1646

\bibitem[\protect\citeauthoryear{{Piatti}}{{Piatti}}{2018}]{p18b}
{Piatti} A.~E.,  2018, preprint, \href
  {http://adsabs.harvard.edu/abs/2018arXiv180908123P} {} (\mn@eprint {arXiv}
  {1809.08123})

\bibitem[\protect\citeauthoryear{{Piatti}}{{Piatti}}{2020}]{piatti2020a}
{Piatti} A.~E.,  2020, \mn@doi [\aap] {10.1051/0004-6361/202039128}, \href
  {https://ui.adsabs.harvard.edu/abs/2020A&A...643A..77P} {643, A77}

\bibitem[\protect\citeauthoryear{{Piatti}}{{Piatti}}{2021}]{piatti2021e}
{Piatti} A.~E.,  2021, \mn@doi [\aj] {10.3847/1538-3881/ac2833}, \href
  {https://ui.adsabs.harvard.edu/abs/2021AJ....162..261P} {162, 261}

\bibitem[\protect\citeauthoryear{{Piatti}}{{Piatti}}{2023}]{piatti2023c}
{Piatti} A.~E.,  2023, \mn@doi [\mnras] {10.1093/mnras/stad2786}, \href
  {https://ui.adsabs.harvard.edu/abs/2023MNRAS.526..391P} {526, 391}

\bibitem[\protect\citeauthoryear{{Piatti} \& {Geisler}}{{Piatti} \&
  {Geisler}}{2013}]{pg13}
{Piatti} A.~E.,  {Geisler} D.,  2013, \mn@doi [\aj]
  {10.1088/0004-6256/145/1/17}, 145, 17

\bibitem[\protect\citeauthoryear{{Piatti} \& {Koch}}{{Piatti} \&
  {Koch}}{2018}]{pk2018}
{Piatti} A.~E.,  {Koch} A.,  2018, \mn@doi [\apj] {10.3847/1538-4357/aadfe1},
  \href {http://adsabs.harvard.edu/abs/2018ApJ...867....8P} {867, 8}

\bibitem[\protect\citeauthoryear{{Piatti} \& {Mackey}}{{Piatti} \&
  {Mackey}}{2018}]{pm2018}
{Piatti} A.~E.,  {Mackey} A.~D.,  2018, \mn@doi [\mnras]
  {10.1093/mnras/sty1048}, \href
  {https://ui.adsabs.harvard.edu/abs/2018MNRAS.478.2164P} {478, 2164}

\bibitem[\protect\citeauthoryear{{Piatti}, {Alfaro}  \&
  {Cantat-Gaudin}}{{Piatti} et~al.}{2019}]{piattietal2019}
{Piatti} A.~E.,  {Alfaro} E.~J.,   {Cantat-Gaudin} T.,  2019, \mn@doi [\mnras]
  {10.1093/mnrasl/sly240}, \href
  {https://ui.adsabs.harvard.edu/abs/2019MNRAS.484L..19P} {484, L19}

\bibitem[\protect\citeauthoryear{{Rain}, {Villanova}, {Mun{\~o}z}  \&
  {Valenzuela-Calderon}}{{Rain} et~al.}{2019}]{rainetal2019}
{Rain} M.~J.,  {Villanova} S.,  {Mun{\~o}z} C.,   {Valenzuela-Calderon} C.,
  2019, \mn@doi [\mnras] {10.1093/mnras/sty3208}, \href
  {https://ui.adsabs.harvard.edu/abs/2019MNRAS.483.1674R} {483, 1674}

\bibitem[\protect\citeauthoryear{{Renzini}, {Marino}  \& {Milone}}{{Renzini}
  et~al.}{2022}]{renzinietal2022}
{Renzini} A.,  {Marino} A.~F.,   {Milone} A.~P.,  2022, \mn@doi [\mnras]
  {10.1093/mnras/stac973}, \href
  {https://ui.adsabs.harvard.edu/abs/2022MNRAS.513.2111R} {513, 2111}

\bibitem[\protect\citeauthoryear{{Rich}, {Shara}, {Fall}  \& {Zurek}}{{Rich}
  et~al.}{2000}]{richetal2000}
{Rich} R.~M.,  {Shara} M.,  {Fall} S.~M.,   {Zurek} D.,  2000, \mn@doi [\aj]
  {10.1086/301156}, \href
  {https://ui.adsabs.harvard.edu/abs/2000AJ....119..197R} {119, 197}

\bibitem[\protect\citeauthoryear{{Ruiz-Lara}, {Gallart}, {Bernard}  \&
  {Cassisi}}{{Ruiz-Lara} et~al.}{2020}]{ruizlaraetal2020}
{Ruiz-Lara} T.,  {Gallart} C.,  {Bernard} E.~J.,   {Cassisi} S.,  2020, \mn@doi
  [Nature Astronomy] {10.1038/s41550-020-1097-0}, \href
  {https://ui.adsabs.harvard.edu/abs/2020NatAs...4..965R} {4, 965}

\bibitem[\protect\citeauthoryear{{Ruiz-Lara}, {Helmi}, {Gallart}, {Surot}  \&
  {Cassisi}}{{Ruiz-Lara} et~al.}{2022}]{ruizlaraetal2022}
{Ruiz-Lara} T.,  {Helmi} A.,  {Gallart} C.,  {Surot} F.,   {Cassisi} S.,  2022,
  \mn@doi [\aap] {10.1051/0004-6361/202244127}, \href
  {https://ui.adsabs.harvard.edu/abs/2022A&A...668L..10R} {668, L10}

\bibitem[\protect\citeauthoryear{{Salgado}, {Da Costa}, {Yong}, {Salinas},
  {Norris}, {Mackey}, {Marino}  \& {Milone}}{{Salgado}
  et~al.}{2022}]{salgadoetal2022}
{Salgado} C.,  {Da Costa} G.~S.,  {Yong} D.,  {Salinas} R.,  {Norris} J.~E.,
  {Mackey} A.~D.,  {Marino} A.~F.,   {Milone} A.~P.,  2022, \mn@doi [\mnras]
  {10.1093/mnras/stac1724}, \href
  {https://ui.adsabs.harvard.edu/abs/2022MNRAS.515.2511S} {515, 2511}

\bibitem[\protect\citeauthoryear{{Saracino} et~al.,}{{Saracino}
  et~al.}{2020}]{saracinoetal2020b}
{Saracino} S.,  et~al., 2020, \mn@doi [\mnras] {10.1093/mnras/staa2748}, \href
  {https://ui.adsabs.harvard.edu/abs/2020MNRAS.498.4472S} {498, 4472}

\bibitem[\protect\citeauthoryear{{Schiappacasse-Ulloa}
  et~al.,}{{Schiappacasse-Ulloa} et~al.}{2025}]{schiappaccaseulloaetal2025}
{Schiappacasse-Ulloa} J.,  et~al., 2025, \mn@doi [arXiv e-prints]
  {10.48550/arXiv.2505.08399}, \href
  {https://ui.adsabs.harvard.edu/abs/2025arXiv250508399S} {p. arXiv:2505.08399}

\bibitem[\protect\citeauthoryear{{Spitoni}, {Matteucci}, {Gratton},
  {Ratcliffe}, {Minchev}  \& {Cescutti}}{{Spitoni}
  et~al.}{2024}]{spitonietal2024}
{Spitoni} E.,  {Matteucci} F.,  {Gratton} R.,  {Ratcliffe} B.,  {Minchev} I.,
  {Cescutti} G.,  2024, \mn@doi [\aap] {10.1051/0004-6361/202450754}, \href
  {https://ui.adsabs.harvard.edu/abs/2024A&A...690A.208S} {690, A208}

\bibitem[\protect\citeauthoryear{{Tsujimoto} \& {Bekki}}{{Tsujimoto} \&
  {Bekki}}{2009}]{tb2009}
{Tsujimoto} T.,  {Bekki} K.,  2009, \mn@doi [\apjl]
  {10.1088/0004-637X/700/2/L69}, \href
  {https://ui.adsabs.harvard.edu/abs/2009ApJ...700L..69T} {700, L69}

\bibitem[\protect\citeauthoryear{{Vanaraj}, {Niederhofer}  \&
  {Goudfrooij}}{{Vanaraj} et~al.}{2021}]{vanarajetal2021}
{Vanaraj} V.,  {Niederhofer} F.,   {Goudfrooij} P.,  2021, \mn@doi [\mnras]
  {10.1093/mnras/stab2094}, \href
  {https://ui.adsabs.harvard.edu/abs/2021MNRAS.507..282V} {507, 282}

\bibitem[\protect\citeauthoryear{{Zhou}, {Kroupa}  \& {Dib}}{{Zhou}
  et~al.}{2025}]{zhouetal2025}
{Zhou} J.~W.,  {Kroupa} P.,   {Dib} S.,  2025, \mn@doi [\mnras]
  {10.1093/mnras/staf1070}, \href
  {https://ui.adsabs.harvard.edu/abs/2025MNRAS.541.1276Z} {541, 1276}

\bibitem[\protect\citeauthoryear{{de Boer}, {Belokurov}  \& {Koposov}}{{de
  Boer} et~al.}{2015}]{deboeretal2015}
{de Boer} T.~J.~L.,  {Belokurov} V.,   {Koposov} S.,  2015, \mn@doi [\mnras]
  {10.1093/mnras/stv946}, \href
  {https://ui.adsabs.harvard.edu/abs/2015MNRAS.451.3489D} {451, 3489}

\makeatother
\end{thebibliography}



\appendix


\begin{table*}
\caption{Properties of star clusters with studies about MPs.}
\label{tab1}
\begin{tiny}
\begin{center}
\begin{tabular}{lccccclccccc}\hline\hline
Name  &  Age  & [Fe/H] & Mass                    & MPs & Ref. & Name &  Age  & [Fe/H] & Mass                     & MPs & Ref. \\  
      & (Gyr) &  (dex) & ($\times$10$^4$$\msun$) &     &      &      & (Gyr) &  (dex) & ($\times$10$^4$$\msun$)  &     & \\\hline
\multicolumn{6}{c}{MW}  &                                         \multicolumn{6}{c}{Helmi}   \\
            &       &        &        &   &   &       &        &        &    &  & \\
NGC~6144	& 13.36	& -1.73	& 8.5	& yes & 5 & NGC~5024    & 12.72 &  -1.97 &  50.2  & yes& 3\\  
NGC~6093	& 13.02	& -1.58	& 32.1	& yes & 5 & NGC~5053    & 12.68 &  -2.23 &   6.3  & yes& 2\\ 
NGC~6218	& 12.97	& -1.26	& 10.6	& yes & 5 & NGC~4590    & 12.17 &  -2.19 &  13.0  & yes& 2\\
NGC~6171	& 12.90	& -0.99	& 6.1 	& yes & 5 & NGC~4147    & 12.13 &  -1.66 &   4.5  & yes& 2\\
NGC~6717	& 12.89	& -1.15	& 2.6 	& yes & 5 & NGC~7492    & 12.00 &  -1.41 &   2.0  & yes& 2\\
NGC~6362	& 12.86	& -1.05	& 11.7	& yes & 5 & NGC~5272    & 11.88 &  -1.48 &  40.9  & yes& 2\\
E~3   		& 12.80	& -0.83	& 2.6 	& no  & 5 & NGC~5634    & 11.84 &  -1.94 &   2.5  & yes  & 21 \\ 
NGC~6723	& 12.77	& -1.02	& 19.7	& yes & 5 & NGC~6584    & 11.75 &  -1.40 &  11.0  & yes& 5 \\
NGC~104		& 12.52	& -0.75	& 85.3	& yes & 5 & NGC~6981    & 11.71 &  -1.40 &   8.1  & yes& 2\\
NGC~6304	& 12.52	& -0.51	& 10.0	& yes & 5 & NGC~6229    & 11.46 &  -1.30 &  21.0  & yes& 4 \\
NGC~6652	& 12.48	& -0.83	& 4.1 	& yes & 5 & NGC~5904    & 11.46 &  -1.25 &  39.2  & yes& 2\\
NGC~6838	& 12.40	& -0.75	& 3.8 	& yes & 5 & NGC~1904    & 11.14 &  -1.37 &  18.0  & yes& 2\\
NGC~6624	& 12.26	& -0.54	& 10.3	& yes & 5 & Ruprecht~106 & 10.85 &  -1.50 &   3.4  & no & 1 \\\cline{7-12}
NGC~6637	& 12.19	& -0.69	& 13.8 	& yes & 5 &          \multicolumn{6}{c}{SMC} \\
NGC~6352	& 12.14	& -0.71	& 6.0 	& yes & 5 &             &       &        &        &     &\\
NGC~6366	& 12.10	& -0.67	& 4.7 	& yes & 5 & NGC~121     & 10.50 &  -1.19 &   37.0 & yes & 6 \\ 
NGC~6388	& 12.03	& -0.77	& 131.0	& yes & 5 & NGC~361     &  8.15 &  -0.90 &   20.0 & yes & 7 \\ 
NGC~5927	& 11.89	& -0.48	& 29.3	& yes & 5 & Lindsay~1   &  7.60 &  -1.04 &   20.0 & yes & 8 \\ 
NGC~6496	& 11.72	& -0.55	& 7.4 	& yes & 5 & Lindsay~8   &  6.45 &  -0.85 &   58.0 & yes & 9\\ 
NGC~6441	& 11.26	& -0.60	& 139.0	& yes & 5 & Lindsay~38  &  6.00 &  -1.39 &   1.5  & no  & 10\\ 
NGC~6791    &  8.00 &  0.30 & 1.9   & no  & 5 & NGC~339     &  6.00 &  -1.15 &    8.0 & yes & 9\\ 
NGC~188     &  6.20 & -0.02 & 0.4   & no  & 5 & NGC~416     &  6.00 &  -0.85 &   16.0 & yes & 9\\ 
Berkeley~39 &  6.00 & -0.21 & 0.4   & no  & 5 & Lindsay~113 &  4.50 &  -1.03 &    2.3 & no  & 11\\ 
Collinder~261 & 6.00& -0.03 & 0.9   & no  & 5 & NGC~419     &  1.30 &  -0.62 &    7.6 & no  & 5 \\\cline{7-12} 
Berkeley~32 & 4.90  & -0.29 & 0.4   & no & 20 &  \multicolumn{6}{c}{LMC} \\
NGC~2682    &  4.50 &  0.05 & 0.3   & no  & 5 &             &       &        &        &     &\\
NGC~2243    & 4.37  & -0.47 & 0.5   & no & 20 &  Hodge~11	& 13.92 & -2.00  & 42.7   & yes & 14 \\ 
Trumpler~5  &  4.27 & -0.35 & 2.4   & no & 20 &  NGC~1841   & 13.77 & -2.02  & 13.2   & yes & 14 \\  
NGC~6253    &  4.00 &  0.40 & 0.4   & no  & 5 &  NGC~1786   & 13.50 & -1.75  & 37.1   & yes & 13 \\ 
Haffner~10  & 3.80  & -0.12 & 0.2   & no & 20 &  NGC~1898   & 13.50 & -1.32  & 77.8   & yes & 13 \\ 
NGC~2425    & 2.40  & -0.14 & 0.3   & no & 20 &  NGC~1466   & 13.38 & -1.90  & 15.8$^{19}$ & yes & 12 \\ 
Berkeley~21 & 2.14  & -0.21 &  0.2  & no & 20 &  NGC~2257   & 12.74 & -1.77  & 7.1$^{19}$ & yes & 14 \\ 
Trumpler~20 & 1.86  & 0.13  & 1.5   & no & 20 &  NGC~2210   & 11.63 & -1.55  & 30.3   & yes & 14 \\ 
NGC~2141    & 1.86  & -0.05 & 1.0   & no & 20 &  NGC~2121   & 3.20  & -0.60  & 10.0   & yes & 5 \\
NGC~2420    & 1.74  &  -0.16& 0.3   & no & 20 &  Hodge~6    & 2.50  & -0.35  &  6.5   & yes & 5 \\
Ruprecht~134 & 1.66  &   0.27& 0.4  & no & 20 &  NGC~1978   & 2.00  & -0.50  & 30.0   & yes & 17 \\
NGC~2158    & 1.55  & -0.16 & 2.1   & no & 20 &  NGC~1651   & 2.00  & -0.30  &  2.7   & no  & 5 \\
NGC~6005    & 1.26  &0.22   & 0.4   & no & 20 &  NGC~1846   & 1.95  & -0.59  & 6.0    & no  & 16 \\
Berkeley~81 & 1.15  & 0.25  &  0.4  & no & 20 &  NGC~2173   & 1.70  & -0.28  & 10.0   & yes & 18 \\
NGC~2477    & 1.12  &0.13   & 1.2   & no & 20 &  NGC~1783   & 1.50  & -0.35  & 26.0   & yes & 15 \\
Trumpler~23 & 0.71  &  0.21 & 0.6   & no & 20 &  NGC~1806   & 1.50  & -0.60  &  1.1   & no  & 5 \\\cline{7-12}
NGC~6802    & 0.66  &   0.15& 0.6   & no & 20 &             &       &        &        &     &    \\
NGC~3532    &  0.40 & -0.02 & 0.5   & no & 20 &             &       &        &        &     &    \\
NGC~6259    & 0.34  & 0.17  & 1.8   & no & 20 &             &       &        &        &     &    \\
NGC~6705    & 0.31  & 0.06  & 1.3   & no & 20 &             &       &        &        &     &    \\
NGC~2516    & 0.24  & -0.05 & 0.4   & no & 20 &             &       &        &        &     &    \\
NGC~6067    & 0.13  & -0.03 & 0.7   & no & 20 &             &       &        &        &     &    \\
Blanco~1    & 0.10  & -0.05 & 0.1   & no & 20\\\cline{1-6}  &       &        &        &     &    \\
\multicolumn{6}{c}{Sagittarius}               &             &       &        &        &     &    \\ 
            &       &        &        &       &             &       &        &        &     &    \\
Terzan~8    & 12.89 &  -2.18 &  7.6   & yes& 1&             &       &        &        &     &    \\
Arp~2       & 11.96 &  -1.66 &  3.9   & no & 1&             &       &        &        &     &    \\ 
NGC~6715    & 11.25 &  -1.34 &  159   & yes& 3&             &       &        &        &     &    \\ 
Palomar~12  & 9.11  &  -0.81 &  0.6   & no & 5&             &       &        &        &     &    \\
Terzan~7    & 7.65  &  -0.58 &  2.4   & no & 1&             &       &        &        &     &    \\
Whiting~1   & 6.50  &  -0.65 &  0.1   & no & 5\\\cline{1-6} &       &        &        &     &    \\
\end{tabular}
\end{center}

\noindent Ref.: (1) \citet{lagioiaetal2025}; (2) \citet{jangetal2025};
(3) \citet{dondoglioetal2025}; (4)
\citet{johnsonetal2017b}; (5) \citet{huangetal2024}; (6) \citet{niederhoferetal2017a};
(7) \citet{p18b}; (8) \citet{niederhoferetal2017b}; (9) \citet{salgadoetal2022}; 
(10) \citet{martocchiaetal2019}; (11) \citet{lietal2019b}; 
(12) \citet{gilliganetal2020}; (13) \citet{vanarajetal2021}; (14) \citet{lietal2025}; 
(15) \citet{cadelanoetal2022}; (16) \citet{ohetal2023}
(17) \citet{saracinoetal2020b}; (18) \citet{kapseetal2022}; (19) \citet{pessevetal2008};
(20) \citet{bragagliaetal2024}: (21) \citet{cb2021}. 
\end{tiny}
\end{table*}

\bsp	
\label{lastpage}
\end{document}